# Detector challenges of the strong-field QED experiment LUXE at the European XFEL

**Yan Benhammou[a]   for the LUXE collaboration**

[a] *Tel Aviv University,*
 *Tel Aviv-Yafo 6997801, Israel*

 *E-mail:* ybenham@cern.ch

The LUXE experiment (Laser Und XFEL Experiment) is a new experiment in planning at DESY Hamburg using the electron beam of the European XFEL (Eu.XFEL). LUXE is intended to study collisions between a high-intensity optical laser and up to 16.5 GeV electrons from the Eu.XFEL electron beam, or, alternatively, high-energy secondary photons. The physics objective of LUXE are processes of Quantum Electrodynamics (QED) at the strong-field frontier, where QED is non-perturbative. The design of the experimental setup and the different detectors are presented.

.

1. Introduction

Quantum Electrodynamics (QED) is a well-tested and well understood theory. The perturbative calculations are precise; however, in the presence of strong fields, perturbative QED breaks down and becomes non-perturbative. Above a critical value of the field (the "Schwinger limit" [1]), the vacuum becomes polarized and there is a production of electron-positron pairs [2].
The LUXE [3] experiment at DESY and the European XFEL (Eu.XFEL) in Hamburg is intended to study strong field QED processes using collisions between a high-intensity laser and a 16.5 GeV electron beam. Depending on the beam parameters, the particle flux will be between $10^{-4}$ to $10^9$; we will present the different sub-detectors and their design to cope with such rates.

2. Physics processes

It is important to introduce two parameters which are relevant in the laser-electron interactions: the first one is the classical non-linearity parameter of the laser field,
$$\xi = \frac{e\varepsilon_L}{m_e \omega_L}$$
where $\omega_L$ is the laser frequency. It can be understood as proportional to the laser intensity. The second parameter is the quantum parameter
$$\chi_e = \frac{E^*}{\varepsilon_{crit}}$$
which, in the case of electron-laser interaction, quantifies the ratio of the effective field strength in the electron rest frame and the critical field. In Figure 1, it is possible to see the potential coverage of the LUXE experiment with respect to past and nowadays experiments.

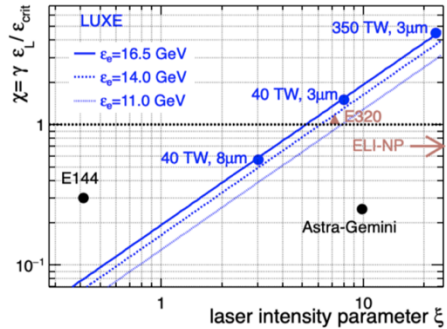

*Figure 1: Potential coverage of LUXE compared to the past experiment (E144) and nowadays experiments (SLAC-E320 (US), Astra Gemini (UK), ELI-NP (RO))*

In the case of electron-laser interactions, the main processes are the non-linear Compton scattering (Figure 2a) and the non-linear Breit-Wheeler pair production process (Figure 2b).

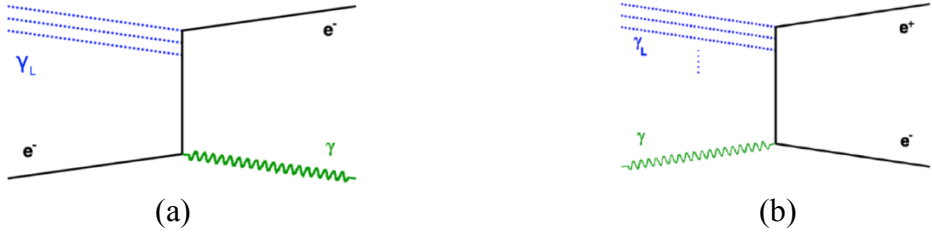

*Figure 2: Non-linear Compton scattering (a) and non-linear Breit-Wheeler pair production (b)*

In Figure 3, we see the rate of photons produced from non-linear Compton scattering as a function of the photon energy for different values of $\xi$. The edges in the photon spectra correspond to the interaction of the electron with several photons of the laser field. The first edge, with the lowest energy appears when the electron interacts with one laser photon, the next one is when the electron interacts with two laser photons and so on. As the laser intensity increases (higher $\xi$), the probability of multiple photons grows. For higher $\xi$, the effective mass of the electron in the field increases, causing the shift of the kinematic edges towards lower energies.

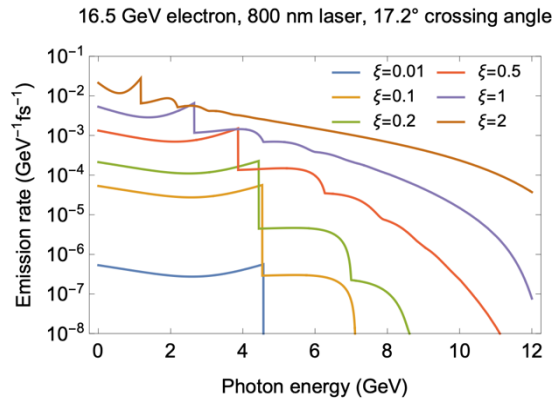

*Figure 3: Photon emission rate as a function of the photon energy*

### 3. Beam setup

There will be two setups for the beam line corresponding to the two different interactions: electron-photon and photon-photon. The LUXE experiment will be located at the end of the Eu.XFEL electron accelerator. A dedicated beam-line was designed to extract one single electron bunch out of about 2500 bunches per Eu.XFEL train, to be guided to the LUXE experimental area at a repetition rate of 10 Hz with an energy of 16.5 GeV [4].

### 3.1 Electron-photon interactions

The laser intended for the LUXE experiment is a femtosecond-pulsed Titanium-Sapphire optical laser ($\lambda L$ = 800 nm) using the chirped-pulse-amplification technique [4]. The laser intensity parameter $\xi$ is varied by changing the laser focal spot size in the interaction point, with the highest $\xi \sim 7.9$ (23.6) in phase-0 (phase-1) achieved with a 3 μ$m$ spot size. The setup is shown in Figure 4.

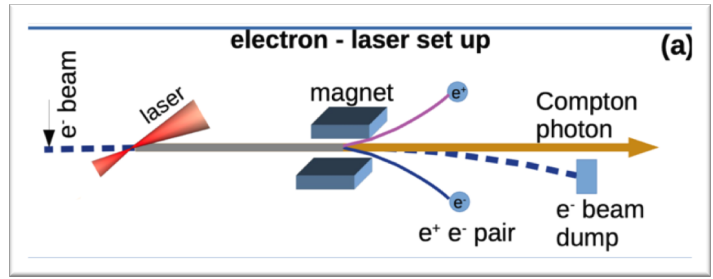

*Figure 4: Setup of the electron laser interaction*

3.2 Photon-photon interactions

To generate some photons, a thin target is placed in the electron beam. The electron beam is deflected by a magnet, and the photon beam continues to the interaction point with the laser, as shown in Figure 5.

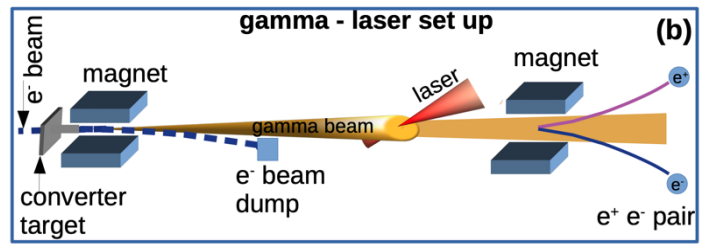

*Figure 5: Setup of the photon laser interaction*

4. The LUXE detector setup

The LUXE detector setup for the electron-laser interaction is shown in Figure 6. Close to the interaction point, the trajectory of the produced electron positron pair is bent by a magnet and on one side, a tracker and an electromagnetic calorimeter are installed for the positron detection. The number of positrons per interaction varies between $10^{-3}$ to $10^5$. On the other side, a scintillating screen and a Cherenkov detector can measure the characteristic of the electrons. There, the number of electrons is between $10^6$ to $10^9$ per bunch crossing.

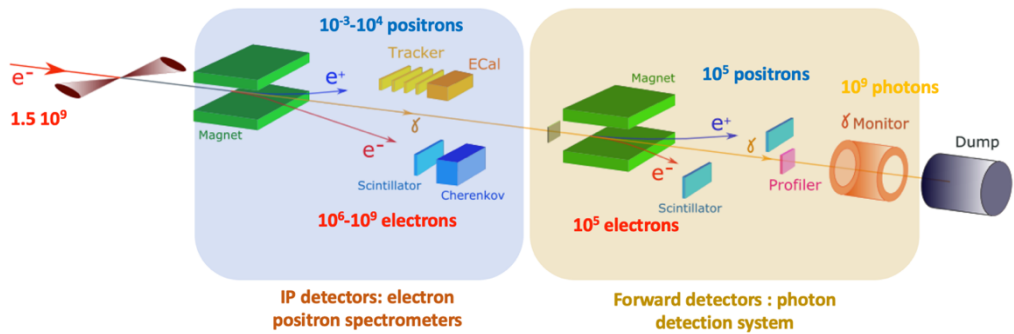

*Figure 6: Experimental setup of the electron-laser interaction.*

The photon beam will interact with a converter (10 μm tungsten) to create electron-positron pairs which will be deflected by a magnet. These pairs will be detected by scintillator screens. The non-interacting photons (~$10^9$) will be detected with two different detectors, then stopped in a dump.

4.1 Positron detection

The positron's trajectory will be bent to a system consisting of a tracker and an electromagnetic calorimeter to measure all their characteristics (Figure 7).

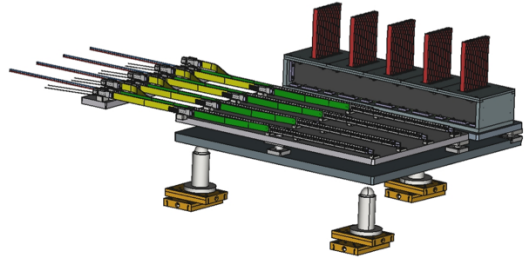

*Figure 7: The positron detector consists of a tracker (green) and a calorimeter (grey). Both systems are on the same precision table*

The number of positrons produced depends on the laser intensity as shown in figure 8; it will vary from few positrons up to $10^5$ (in phase 1 with ξ ~ 7).

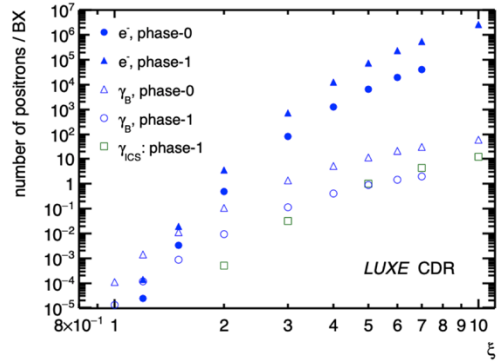

*Figure 8: Number of positrons per bunch crossing as a function of the laser intensity*

The tracker is composed of three layers of silicon pixel detectors (ALPIDE) developed by the ALICE collaboration [6]. The pixel size is 27x29 μm² which allows to reach a position resolution around 5 μm. It has a very high tracking efficiency (>99 %) and a low material budget (X/$X_0$ = 0.357%).

The electromagnetic calorimeter is composed of 20 layers of 3.5 mm tungsten (1 $X_0$) separated by one millimeter. In between, a sensor with 5x5 mm² pads is detecting the electromagnetic shower. Two types of sensors are under study, silicon and gallium arsenide. It has been designed for the ILC forward region. A dedicated ASIC, the FLAME, has also been developed for the ILC [7]. For single electrons hitting the calorimeter perpendicular to the front face and after reconstruction, the energy and position resolution are measured to be $\frac{\sigma(E)}{E} = \frac{19.3\%}{\sqrt{E}}$ and $\sigma(x) = 780 \, \mu m$.

4.2 Electron detection

The electrons detection is challenging because of their rate that can reach up to $10^9$ in the electron-laser mode. The detection system consists of a scintillator screen and a Cherenkov detector (Figure 9).

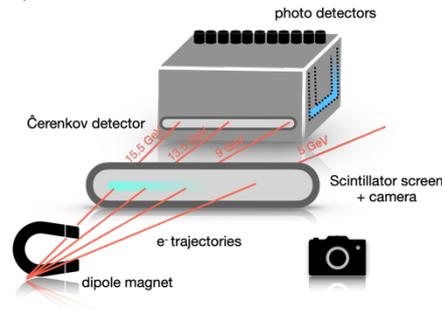

*Figure 9: Schematic of the electron detection system. A camera is taking pictures of the scintillation light and a Cherenkov detector measures the energy of the electrons*

The scintillator screens have been developed by the AWAKE collaboration at CERN [8]. A camera takes a picture of the scintillation light and can reach a position resolution around 500 μm. This allows a separation between background and signal around 100. These screens are radiation hard up to 100 MGy.

The Cherenkov detector is filled with argon and it measures the particle rate as a function of the deflection by the magnetic field. It helps to reject photons and low-energy backgrounds due to the Cherenkov threshold of ~20 MeV.

4.3 The photon detectors.

After the convertor target, the rates will be around $10^4$-$10^5$ for electrons/positrons and $10^9$ for the photons. Three technologies are used:
a spectrometer composed of scintillator screens coupled with photo-cameras for the electron/positron detection, a gamma profiler to measure the angular spectrum of photons, and a photon monitor. The gamma profiler will be composed of two sapphire (very hard radiation material, up to 100 MGy) strip detectors perpendicular to the beam located on a micron precision xy table [9]. Each sensor is a 2x2 $cm^2$ sapphire sensors 100 μm thick, with 100 μm strips. The precision will be about 5 μm in x and y.
The photon monitor will allow to measure the flux of particles back scattered from the dump. It consists of eight lead glass blocks located 10 cm upstream the photon beam dump, around the beam axis with a radius of 17cm. Photomultiplier tubes are reading the light for these glass blocks. There is almost a linear dependence between the energy deposited and the number of incident photons, with an uncertainty between 3% and 10%.

5.Conclusion

The LUXE experiment will explore the strong-field QED predictions using the European XFEL and a high-power laser. All the detectors have been tuned to cope with the large dynamic range of particle rates, from $10^{-2}$ to $10^9$ per bunch crossing.